\documentclass[12pt,preprint]{aastex}

\def\sun{\hbox{$\odot$}}

\def\pasa{\ref@jnl{PASA}}               % Publications of the ASA

\lefthead{}
\righthead{}

\begin{document}
\title{A Search for Stellar Populations in High Velocity Clouds}

\author{M. H. Siegel\altaffilmark{1}, S. R. Majewski\altaffilmark{2,3}, 
C. Gallart\altaffilmark{4}, S. Sohn\altaffilmark{2,5}, W. E. Kunkel\altaffilmark{6}, R. Braun\altaffilmark{7}}

\altaffiltext{1}{University of Texas -- McDonald Observatory, Austin, TX, 78712 
(siegel@astro.as.utexas.edu).}
\altaffiltext{2}{University of Virginia, P.O. Box 3818, Charlottesville, VA, 
22903 (srm4n@virginia.edu,ss5fb@virginia.edu).}
\altaffiltext{3}{David and Lucile Packard Foundation Fellow.}
\altaffiltext{4}{Ram\'on y Cajal Fellow, Instituto de Astrof\'\i sica de Canarias
38205 La Laguna, Tenerife, Spain, (carme@iac.es).}
\altaffiltext{5}{Present Address:  
Korea Astronomy and Space Science Institute, 61-1, Whaam-Dong,
Youseong-Gu, Daejeon 305-348, Korea (tonysohn@kao.re.kr).}
\altaffiltext{6}{Las Campanas Observatory, Casilla 601, Las Serena, Chile (kunkel@jeito.lco.cl).}
\altaffiltext{7}{Netherlands Foundation for Research in Astronomy, 
PO Box 2, 7990 AA Dwingeloo, The Netherlands (rbraun@nfra.nl).}

% version date December 30, 2004

\begin{abstract}
We report the results of a photometric search for giant stars associated with the cores of four 
high velocity clouds (HVCs) -- two of which are compact HVCs -- using
the Las Campanas Du Pont 2.5 meter and Cerro Tololo Blanco 4 meter telescopes in combination with a 
system of filters (Washington $M$, $T_2$ + $DDO51$) useful for identifying low 
surface gravity, evolved stars.  Identical observations of nearby control
fields provide a measure of the ``giant star'' background.
Our data reach $M_0=22$ for three of the HVCs and $M_0=21.25$ for the fourth, depths that
allow the detection of any giant stars within 600 kpc.  
Although we identify a number of faint late-type giant star candidates, we find neither a coherent red giant branch 
structure nor a clear excess of giant candidate counts in any HVC.
This indicates that the giant candidates are probably not related to the HVCs and are
more likely to be either random Milky Way giant stars or one of several classes of potential survey
contaminants.  Echelle spectroscopy of the brightest giant 
candidates in one HVC and its control field reveal radial velocities representative of the canonical Galactic stellar
populations.  In addition to these null results, no evidence
of any young HVC stellar populations --- represented by blue main sequence stars --- is found, 
a result consistent with previous searches.  Our methodology, specifically designed to find faint 
diffuse stellar populations, places the tightest upper limit yet on
the {\it total} stellar mass of HVCs of a few $10^5 M_{\sun}$.
\end{abstract}

\section{Introduction}

The high velocity clouds (HVCs) are a population of neutral hydrogen clouds with LSR velocities of 80-400 
km s$^{-1}$ (Muller et al. 1963). Despite nearly four decades of study, the nature of the clouds is 
still a matter of some debate (see, e.g., Wakker \& van Woerden 1997; Putman \& Gibson 1999) at least partly
because of the difficulty in measuring distances to HVCs.  Without distances, the nature of the HVCs --
e.g., whether they are a Galactic or extragalactic phenomenon -- 
can only be ascertained by indirect means.

It is clear that the dominant HVC feature -- the 100$^{\circ}$ long Magellanic Stream -- is the result of
interaction between the Magellanic Clouds and the Milky Way, either as a result of tidal disruption or ram-pressure 
stripping or both (Wannier \& Wrixon 1972; Matthewson, Clearly \& Murray 
1974; Putman et al. 1998, 2003; Putman \& Gibson 1999).  
Some HVCs, particularly Complex C, have been attributed to
gas falling onto the Milky Way from similar but perhaps smaller-scale interactions (Larson 1972; Tosi 1988; 
Gibson et al. 2001; 
Tripp et al. 2003) while other HVCs are ascribed as the product of Galactic fountains 
(Shapiro \& Field 1976; Bregman 1980, 1996).

The compact high velocity clouds (CHVCs) are possibly the most contentious subset of HVCs.  It has 
been proposed 
that the CHVCs could be the ``missing satellites'' of the Local Group
(see discussion in Braun \& Burton 1999 and 2000, hereafter BB99 and BB00; Blitz et al. 1999; de 
Heij et al. 2002a; Maloney \& Putman 2002; Maller \& Bullock 2004), whose existence
is presumed on the basis of their formation in N-body simulations of structure formation in the presence of
Cold Dark Matter (CDM).  These simulations predict the mass distribution of the Local Group 
to follow a power law 
$N(M) \propto M^{-\alpha}$ where $\alpha$ is at least 2 (see, e.g., Klypin et al. 1999; Moore et al. 1999).  
Such a spectrum would imply the existence of several hundred low mass dwarfs in the Local Group, an order of 
magnitude more than detected, despite a number of optical surveys for
low surface brightness members (Mateo 1998; Armandroff et al. 1998, 1999; 
Karachentseva \& Karachentsev 1998, 2000; Karachentsev \& Karachentseva 1999; Karachentsev et al. 2000; 
Whiting et al. 2002; Willman et al. 2004).  
This contradiction would resolve if the Local Group were filled with dark galaxies -- objects 
containing dark matter but few, if any, stars, perhaps because of inhibition of star formation 
in the early Universe (see, e.g., Bullock et al. 2000). 

The CHVCs' kinematical preference for the Local Group 
standard of rest (BB99) and the existence of a small overdensity of 
clouds toward M31 with high negative velocities (Burton et al. 2002; de Heij et al. 2002a) 
suggest Local Group scale distances.  High resolution 
mapping (BB00; Burton et al. 2002) shows several CHVCs to have core-halo morphology, similar to that expected 
for dark matter-dominated mini-halos.  Three independent methods yield distance estimates for some HVCs of 
$\sim$ 0.2-0.7 Mpc (BB99; BB00; Burton et al. 2001; Robishaw et al. 2003).  The CHVCs also show velocity 
gradients along their major axes with implied rotation speeds of $V_c \sim 15-20$ km s$^{-1}$ (BB99; BB00; 
Burton et al. 2001) as well as high internal velocity dispersions.  Both phenomena imply high dark matter 
content (BB00), assuming the HVCs are virialized.

Conversely, the sky distribution of CHVCs appears
inconsistent with a Local Group origin (Putman et al. 2002; cf. B99 and de Heij et al. 2002a) and CHVCs are not
definitively seen in extragalactic groups (de Blok et al. 2002; Zwann 2001; 
Banks et al. 1999; Pisano et al. 2004; cf. Braun \& Burton 2001).  Maloney \& Putman (2003) evaluated the effect of 
extragalactic ionizing radiation on the CHVCs and concluded that the properties of CHVCs are more consistent with a 
Galactic halo origin than a cosmological one.  And while anomalous hydrogen clouds have been found in nearby
galaxies (M31; Thilker et al. 2004: M53 and M81; Miller \& Bregman 2004), these clouds are closer to their 
parent galaxies than would be likely for dark galaxies.

One puzzling aspect of CHVCs are the chemical abundances.  They are about a tenth solar, a value too 
low for Galactic fountain models (Wakker et al. 1999; BB99; BB00) but too high for protogalactic fragments,
given the lack of recent star formation in HVCs (Blitz et al. 1999; BB99; BB00).  It might be consistent, however, 
with a tidal or ram-pressure stripping origin since many Local Group dwarfs have abundances near this range.

The detection of stars in HVCs would open a new dimension to their study and quickly resolve many uncertainties,
including their distance and chemical enrichment history.  Additionally, if the CHVCs were discovered to have
stellar populations while the HVCs did not, this would clearly indicate that different mechanisms were 
responsible for the formation of the compact and non-compact HVCs.

If the HVCs were within one Mpc of the Milky Way, they would have 
red giant stars bright enough to be detected with modest aperture
telescopes.  A number of recent studies have searched for HVC stars.  Willman et al. (2002) surveyed a 
small number of HVCs 
that were included in the Sloan Digital Sky Survey and found no indication of stellar content.  
Hopp et al. (2003) surveyed only CHVCs, combining deep VLT imaging
with data from the 2MASS survey and comparing
each CHVC to a nearby control field.  The $VI$ data rule out significant stellar content to a 
distance of 2 Mpc, although they cover small regions of each cloud.  The 2MASS data cover much 
larger areas of sky, but would only detect giant stars to a distance of 300 kpc.
The most complete survey, by Simon \& Blitz (2002), also focused on CHVCs 
and failed to detect any stellar content.  This study placed a lower limit of 100 kpc 
on the distance to 250 CHVCs and even greater lower limits (300-1400 kpc) on the distance to 60 CHVCs.  The 
Simon \& Blitz survey searched POSS 
plates for enhancements of stellar density at the positions of the CHVCs and followed up potential detections with 
monochromatic observations in the Spinrad $R_s$ filter.  Though impressive in scale, the Simon \& Blitz survey was
biased toward larger HVC stellar populations with densities that could rise above 
the statistical noise of the foreground of Milky Way stars.  
Diffuse, low surface brightness systems are a much greater challenge to identify
against the Poissonian noise of dense foreground/background Galactic starcounts.  To find such structures
requires observational strategies that increase their detectability. 

In this article, we present the first search for {\it low surface-brightness} stellar populations in HVCs, a search
based on the observational strategy of minimizing the effects of Galactic contamination.
Our photometric survey of four HVCs and adjacent control fields utilizes a filter
system especially sensitive to red giant/supergiant stars --- the primary stellar type sought in
distant HVCs (and a stellar type found in stellar populations of all ages and metallicities) 
--- and that allows a large fraction of Galactic stellar contaminants --- which will be 
dominated by main sequence stars --- to be identified and eliminated.  The substantially
reduced background/foreground makes it possible to look for very tenuous CHVC stellar populations
that would be swamped within simple starcount analyses.
For one of our target HVCs and control fields we supplement our photometric analysis with
radial velocities derived from echelle 
spectroscopy of the brightest giant star candidates.   \S2 of this paper details 
the observation and reduction of our photometric observations.  \S3 presents an
analysis of these data and attempts to quantify the residual sources of possible ``noise'' 
in our program after elimination of the bulk of the likely Galactic dwarf stars.
\S4 details the spectroscopic work on the brighter giant candidates in one of the HVCs, work that 
confirms the basic analysis of \S3.  \S5 defines the limits that our survey places on the distances
of any stellar populations in the HVCs. 

\section{Observations and Reduction}

The high velocity clouds HVC 017-25-218, HVC 030-51-119 and HVC 271+29+181 were observed
in the Washington $C$, $M$, and $T_2$ filters as well
as the MgH+Mgb triplet sensitive $DDO51$ filter on UT 5-8 June 2000.  We used the Du Pont 2.5 meter telescope and the 
Wide Field Camera (WFC) designed by Ray Weymann and collaborators.  The WFC has a circular field of view of
23' at a pixel scale of $\sim$0\farcs7 per pixel.  The optical defect in the WFC noted by Majewski et al. 
(2000a) was corrected prior to the observations, providing a useable, but 
still modest quality point spread function (PSF) over the entire image (see below).  
Each HVC was paired with a control field displaced
by $\sim2$-$3$ degrees at nearly constant Galactic latitude
and observations were switched back and forth between the HVC and control, 
resulting in identical depth and image quality for each target and paired
control field.  Total exposure times for each HVC and its control field were 11500-13400 seconds in 
$DDO51$, 10000-11000 seconds in $C$, 1830-2130 seconds in $T_2$ and 2730-2790 seconds in $M$, integration times
that provide 
completeness in all four passbands to beyond $M=22$.  HVC 030-51-119 and its control 
field were only minimally observed in $C$.

We observed the high velocity cloud HVC 267+26+216 with the Blanco 4-m
Telescope and Mosaic II
8k$\times$8k CCD at CTIO during the nights of UT 26-27 February, 2000.  The CCD array has a pixel scale of $0\farcs27$
per pixel resulting in a field of view of $37\arcmin$.  The HVC and control field
were observed with total exposure times of 300 seconds in $M$ and $T_2$, and 2100 seconds in $DDO51$.  No
data were obtained in $C$ for this HVC.

Detailed radio properties of the observed HVCs are listed in B99.  Figure 1 shows HIPASS (Barnes
et al. 2001, Putman et al. 2002) radio images of the four target HVCs 
with the test and control fields marked.  Coordinates and average reddening coefficients (from
Schlegel et al. 1998) for the four HVCs and their control fields are listed in Table 1. 
We note that the control fields for HVC 017-25-218 and HVC 030-51-119 are not ideally placed.  They were 
positioned before HIPASS (Barnes et al. 2001, Putman et al. 2002) maps became available and unfortunately landed 
on the extended structure of the target HVCs.  However, the locations of the control fields have much lower 
column densities than the central fields and should still provide a useful contrast to the cores.  Analysis
of the HIPASS data indicates that HVC 017-25-218 and HVC 271+29+181 meet the definition of {\it compact}
high velocity clouds.\footnote[1]{At the time of our optical observations, HVC 030-51-119 and HVC 267+26+216 
were classified as CHVCs.  Later radio observations by HIPASS revealed a less compact structure 
than previous radio images, so that these two objects would no longer be considered CHVCs.  In 
this paper we have reclassified these two objects as HVCs.}

Reductions were carried out through the standard IRAF\footnote[2]{IRAF is distributed by the 
National Optical Astronomy Observatories, which are operated by the Association of Universities for 
Research in Astronomy, Inc., under cooperative agreement with the National
Science Foundation.} CCDRED and MSCRED pipelines as appropriate to each data set.  
The WFC produces a roughly circular image on the square CCD chip.  The unexposed areas of the chip
consist entirely of CCD noise and can render normal pipeline analysis difficult.
We produced an image mask from the flat fields to remove these sections from analysis.  
The location of the unexposed region of the chip, however, varied 
slightly over the course of the run.  As a result, the unexposed regions occasionally extended
beyond the nominal mask.  These ``noisy edges'' were generally small ($\leq 0.5\% 
$ of the total chip area) but
produced hundreds of ``detections'' during photometry.  The detections from the noisy edges were 
clearly non-stellar and were easily removed by our stringent morphological classification 
(see \S3).  A re-analysis of our data with more restrictive masking that excludes more of the field 
edges showed no change in the starcount density.

We derived photometric measures with the DAOPHOT/ALLFRAME PSF photometry package (Stetson 1987, 1994).  
The PSFs were fit with a quadratically varying Moffat function.  The Du Pont data 
produce modest quality PSFs, with typical rms residuals of 5-7\% per pixel.  The Blanco images 
exhibit excellent PSFs, with typical rms residuals of 1-2\% per pixel.

All photometry is calibrated to the standard stars of Geisler (1990, 1996) using the matrix
inversion and iterative techniques to derive and apply transformation equations described in 
Siegel et al. (2002).  For the Du Pont data, both 
program and standard stars are measured in identical $3 \times FWHM$ circular apertures.  
The Blanco MOSAIC data are calibrated using curve-of-growth analysis from DAOGROW 
(Stetson 1990).  The uncertainty in the photometric transformations is approximately 
0.01 magnitudes in all filters.

We derived astrometry for our stars from the reference frame of the USNO-A2.0 catalogue 
(Monet et al. 1996) using the STSDAS TFINDER program in IRAF.  Reddening values were generated for each 
individual star based on interpolation of the reddening maps of Schlegel et al. (1998).

Final photometric catalogues are available in electronic form from the Astrophysical Journal.  The
catalogues list the photometric measures for stars (as selected in \S3) and include $(\alpha, \delta)
_{J2000.0}$, photometric measures and $E_{B-V}$ reddening measures.  We have also included measures of a 
modified Welch-Stetson (1993) variability index in which non-variable stars would have values $\leq1.0$.  
The 4-meter data only have one observation in each filter and the variability indices are listed as 0.0.
The stars from the 2.5-meter data generally have indices $\leq 1.0$ indicating little variability over the three
days of our observing program and agreement between the formal errors and empirical scatter of the 
photometric measures.  Bright stars are occasionally 
saturated on the CCD images, resulting in elevated variability indices.

\section{Analysis}

The $M$, $T_2$, $DDO51$ filter system is now well-established for its ability to distinguish late-type 
high surface gravity dwarf stars from late-type, low surface gravity, evolved giant stars through the use 
of two color diagrams (Majewski et al. 2000a,b; Siegel et al. 
2000; Palma et al. 2003; Sohn et al. in prep.; Westfall et al. in prep).  In particular, we reference Figure 5 of 
Palma et al. (2003) and Figure 5 of Majewski et al. (2000a), which demonstrate 
expectations for the appearance of a red giant branch (RGB) in the two-color diagram.  

Our data reach a magnitude or more deeper than $M\sim 22$.  However, 
data fainter than $M\sim22$ have larger uncertainties and less reliable morphological classification. We can reliably
classify objects brighter than $M_0=22.0$ for the Du Pont data and $M_0=21.25$ for the Mayall data
\footnote[3]{We classify stars based on the DAOPHOT SHARP
parameter.  For the 2.5-meter data, objects with $-0.5 < SHARP < 0.11$ are classified as stars; for the 4-meter
data, the criterion was $-0.2<SHARP<0.11$.  See Siegel et al. (2002) for a discussion of the efficacy of 
DAOPHOT parameters for object classification.}.  Figure 2 shows the run of photometric error with magnitude for our 
data.  We apply
magnitude error limits to remove obviously poor measures but retain the bulk of the stars.  The adopted magnitude 
error limits 
are $\sigma_{T_2}$, $\sigma_M$ and $\sigma_{DDO51} <$ 0.06 for the Du Pont data and $\sigma_{T_2}$, $\sigma_M$ and 
$\sigma_{DDO51} <$ 0.035 for the Mayall data.

Figure 3 shows color magnitude diagrams (CMDs) of our four HVCs and their corresponding
control fields.  Because they are at similar Galactic latitudes, are dominated by Milky Way stars, and have identical
quality data, the HVC and control field CMDs look similar.

The CMDs clearly show
that the HVCs fields do not contain young stars (at least if they are closer than a few hundred kpc, a limit
that depends on the age of the main sequence turnoff [MSTO]) since no obvious main sequences 
project blueward of the ``blue edge'' of the old, metal-poor halo stars at $(M-T_2)\sim 0.75$.
(The slight blueward shift of the blue edge at faint magnitudes is a result of the transition to the 
more metal-poor populations of the inner halo and thick disk.)
This is not altogether surprising, since the previous optical searches of 
HVCs (B99; B00; Simon \& Blitz 2002) have already placed tighter constraints 
on the absence of young stars in HVCs than our data by virtue of the greater number of clouds surveyed, 
deeper photometry and/or larger spatial area.

The principal advantage of our program is the degree to which we can detect old or 
intermediate age stellar populations --- particularly very diffuse ones.  
Such populations would have RGBs well redward of the
blue edge.  Ordinarily, with simple starcount techniques, a weak RGB might easily be swamped by 
the foreground of red Galactic dwarf stars and 
fall below the detection threshold of the previous surveys.  Our program overcomes this problem by 
severely reducing the Galactic foreground.

The basic premise of the three-filter selection method is that late-type giants and dwarfs occupy different
loci in color-color space (Fig.\ 4).  The dwarf stars lie along the arc of stars at negative $M-DDO51$
that traces the strength of magnesium absorption as a function of effective temperature in high 
surface-gravity, main sequence stars.  Low surface-gravity giants lie above this curve in general, offset to a 
degree dependent upon metallicity (Majewski et al. 2000b).

We interactively fit a sixth order polynomial to $(M-DDO51)_0$ as a function of $(M-T_2)_0$ over the 
limit $0.7 < (M-T_2)_0 < 3.0$.  We gradually removed outliers from this function so that it eventually was
entirely defined by stars along the dwarf sequence.  Once this 
system was defined, we derived each star's distance from the nominal dwarf function as:

\begin{center}
$\Delta = (M-DDO51)_0 - f (M-T_2)_0$
\end{center}

\noindent where $f(M-T_2)_0$ is the ``best-fitting" locus to the dwarf stars in the two-color diagram.
We then flagged any star
with $1.0 < (M-T_2)_0 < 2.5$ and a positive $\Delta$
value greater than three times its photometric error 
($\sigma = \sqrt{\sigma_M^2 + \sigma_{T_2}^2 + \sigma_{DDO51}^2}$) as a potential giant.
\footnote[4]{Since the shape of the dwarf star curve will
change with age/abundance (see Paltoglou \& Bell 1984), 
using a single fit for all of the dwarf stars is too simplistic.
To take out first order effects in the variation of the dwarf locus, 
we actually measure $\Delta$ with respect to the locus as defined by
stars within one magnitude of each star's $M$. }

This approach provides some distinct advantages over previous efforts that used rigidly defined selection
regions in color-color space 
to select giant candidates 
(see Majewski et al. 2000a,b; Morrison et al. 2001; Palma et al. 2003).  First, if the errors are Gaussian, each 
giant candidate selected by the above method 
is 99.7\% likely to have photometry placing it genuinely outside the nominal dwarf locus.  Second,
we obtain an immediate analytical estimate for the contamination level as 0.3\% of the total number of stars.  
Third, even if the errors are not Gaussian or are underestimated, the 
number of stars with {\it negative} $\Delta$ values greater than $3\sigma$ gives a second, robust estimate of the 
photometric contamination level.  Finally, by removing the constraints of rigidly defined selection limits, a star 
with exceptional photometry 
is flagged as a giant even if only a short distance from the nominal dwarf locus while a 
star with poor photometry must pass correspondingly more stringent requirements to be selected as a potential giant.

Figure 5 shows the $\Delta$ versus $M$ distribution of the stars in our eight HVC and control fields.  The trend is 
roughly linear, with growing dispersion as the faint end.  Stars flagged as giants and negative outliers are marked.  
Figure 6 shows the color-color diagram of our giant and negative outlier stars.  Note that the reddest giant candidates 
are nearly within the apparent dwarf locus.  These are stars with exceptionally small photometric uncertainties.

Table 2 lists the number of giant candidates in each HVC and its corresponding control field, while Figure
7 shows the CMDs of the giant star candidates in the HVCs.  The advantage of our method becomes immediately
obvious from comparing the raw starcounts to giant star candidate counts in Table 2 or by comparing Figure 7 
to Figure 3.  The screening for giant star candidates
reduces the foreground dramatically -- by more than an order of magnitude in all eight
cases.  A $3\sigma$ surface brightness enhancement would have to be 60-200 stars strong to show up in the 
raw starcounts or uncleaned CMD.  However, Figure 7 and the last columns of Table 2 would easily reveal the signature
of even a handful of HVC RGB stars, with a $3\sigma$ surface brightness enhancement corresponding to only 5-25 stars, 
an order of magnitude improvement in sensitivity.

The HVC fields have a significant number of giant candidates that, on first blush, would appear to show
a stellar population in the HVCs.  However, the control fields provide a critical second constraint.  
The HVC fields show no excess of giant candidates over the
control fields.  In fact, three of the HVC fields have {\it fewer} giant candidates than their control fields.

The color-magnitude distribution of giant candidates in the HVCs and their corresponding control
fields are similar, although they show a number of interesting features.
Most of the giant candidates are toward the fainter end of the data, where
contamination by extragalactic objects and photometric errors is most likely.  However, CHVC 017-25-218 and its
control field also show a diffuse clump of bright stars around $M_0\sim15-17$.  If bona fide giants, these
bright stars would be at distances of 40-100 kpc and therefore well within the bounds of the
outer Milky Way halo.

There are few giant candidates with colors near the tip of the red giant branch (TRGB) -- typically found
at colors
of $1.8 \leq M-T_2 \leq 2.5$ for the metallicity range $-2.0 \leq [Fe/H] \leq -0.5$ (see Table 5) and no
obvious RGB in any of our fields.  The only
potential TRGB stars are in the control field of HVC 271+29+181.  This field shows a clump of stars near 
($M_0, (M-T_2)_0) \sim (21.5, 1.9$) as well as a sprinkling of brighter red stars.  The latter we have found
to be field star contaminants (see \S4) while the latter could be a very 
distant ($> 450$ kpc), very dispersed RGB tip.

In any case, it is clear that the HVCs are not host to any significant stellar population --- and our discussions
and analysis below suggest that they are not likely even to host {\it meager} stellar populations.
The most likely explanation for the presence of ``giant star candidates" in our fields is that 
these candidates are either field giants from the Galactic halo, or represent contamination of the
giant sample by non-giants.
We now quantify these contributions and in \S4 show that 
the radial velocity distribution in one of our fields is consistent with that of
Galactic field stars.

\subsection{Photometric Errors}

As discussed above, our new method of giant star selection should provide an immediate evaluation of the photometric
contamination of the giant star region in the two-color diagram.  
Table 3 lists, for each field, the number of contaminants expected given the
3$\sigma$ selection criterion and the number of stars with negative $\Delta$ values greater than 3$\sigma$.
For several fields, the latter number is significantly greater than the former.  This is a result
of the modest quality of the WFC PSF.  A disproportionate number of the negative $\Delta$ stars are located near 
the chip edges where the quadratic PSF is not quite adequate.
We use the greater of the two numbers to represent the level
of photometric contamination, thus providing a conservative estimate for the number of 
{\it bona fide} giant stars within our candidate sample.

\subsection{Subdwarf Contamination}

Another potential source of sample contamination is subdwarf stars.  
Our candidate selection is based on the presumption that the observed 
weaker magnesium absorption in these stars is due to surface gravity effects.
However, very metal-poor ($[Fe/H] \leq -2.0$) halo subdwarfs may contain so little magnesium that their
absorption lines are intrinsically weak and therefore insensitive to surface gravity.  Such stars would also
fall into the giant
region of the two-color diagram.  Morrison et al. (2001) estimated this contamination level
to be small at bright magnitudes but to grow rapidly at faint magnitudes, depending on the $(l,b)$ of
the field.  Using 
their Figure 2, we estimate that our fields should each contain approximately ten subdwarfs, with the
exception of HVC 017-25-218, which should contain approximately 20 subdwarfs.  This could provide a substantial
fraction of the excess ``giant candidates'' in Table 2.

\subsection{Compact Galaxies}

The light from distant galaxies has a significant, perhaps even dominant 
contribution, from giant stars (see, e.g. Bershady 1994 and references therein).   In addition, at high enough
redshifts, the magnesium features can shift out of the $DDO51$ passband, resulting in bluer $M-DDO51$ colors.
An examination of the non-compact objects in our photometry shows that they primarily occupy the giant region, 
a result we have seen in other deep photometry samples that use this filter system.
Our structural parameters should limit galaxy contamination. However,
{\it compact, star-like} extragalactic objects, which are 
expected to contaminate photometric samples at a level of $\sim200$ degree$^{-2}$ to $V\sim21.5$ 
(Kron et al. 1991), could remain in the sample.  Of course, only a small fraction of compact 
galaxies have colors red enough to fall into our selection region in the two-color diagram.

To estimate the contribution of compact galaxies to the giant star selection region, we took quasar counts
from a recent analysis of the Sloan Digital Sky Survey (Richards et al. 2001).  We 
transformed the $M-T_2$ color limits of our giant selection region ($1.0 < (M-T_2)_0 < 2.5$) to approximate $g-r$ 
colors ($0.4 < g-r < 1.5$) using transformations in Majewski et al. (2000b) and Fukugita et al. (1996).  
Approximately 10-15\% of the Richards et al. sample fall into 
this color range resulting in a compact galaxy contamination level of $20-30$ degree$^{-2}$ in 
each of our fields, which would be three galaxies for each WFC field
and nine for the MOSAIC fields.

This estimate would probably be a lower bound for the 2.5-meter data, which reaches a depth
of $M_0=22.0$ and may suffer from more extragalactic contamination owing to the mediocre
PSF and commensurately lower quality morphological classification.  On the other hand, it is probably an upper bound 
for the 4-meter data given its brighter magnitude limit and superior imaging capability.  We choose 
to split the difference and apply a uniform compact galaxy contamination level of five objects per field.

The $C$ band photometry in three of our HVCs provides an observational test of this approximation.
These data were obtained because $C-M$ colors 
would allow us, had we found a coherent RGB in any of our fields, to place constraints on the 
metallicity of the giant stars using the 
iso-metallicity lines derived by Geisler et al. (1991).  
However, the two-color diagram (Figure 8) shows something 
unexpected.  Many of the giant candidates have far bluer $C-M$ colors than even very metal-poor populations.
We find that many of the non-compact galaxies in our photometry also occupy this region of color-color space.
Moreover, the star-like objects most displaced from the iso-metallicity lines are
preferentially nearer to the faint limit of the data -- where galaxy contamination is
expected to be strongest. 

Our interpretation is that some of our faintest giant star candidates are compact galaxies and fall into
the giant region as a result of their composite spectral energy distribution being dominated by giants, or 
by being at high enough redshift to shift their magnesium features out of the $DDO51$ band.
The giant candidates that fall along the iso-metallicity lines in figure 8
are far more likely to be bona fide giants or subdwarfs.  Significantly, the number of giant candidates
substantially removed 
from the iso-metallicity lines appears to be approximately 1-10 per field, consistent with
the number of extragalactic contaminants estimated above.

\subsection{Foreground Giants}

As mentioned above, the brighter giant candidates are likely from the Galaxy itself.  We estimate 
the number of Galactic field giants 
for each field by numerically integrating the Fundamental Equation of Stellar Statistics (von Seeliger 1898) using the 
density laws derived by Siegel et al. (2002, their table 6), and color-absolute magnitude and giant star luminosity 
functions from 
Bergbusch \& Vandenberg (2001) converted to Washington colors using the formulae in Majewski et al. (2000b).  The 
number of total estimated contributed Galactic giant stars is listed in the fourth column of Table 3.  This includes 
contributions from the inner halo, thick disk and bulge.

Column five of table 3 lists the total combined predicted contamination level by subdwarfs, compact galaxies, 
photometric errors as well as the contribution of foreground Milky Way giants.  
Column six lists the difference between the predicted contamination 
level and the actual number of giant candidates.
Although these numbers come with substantial uncertainties, the comparison shows 
some patterns.  HVC 030-51-119 and HVC 267+26+216 
appear to have fewer giant candidates than predicted; their 
control fields match the predictions almost exactly.  On the other hand, the two compact HVCs 
have a clear excess of giant candidates beyond the expected contamination level, an excess matched in their 
control fields.  These discrepancies are larger than the statistical noise of the sample.  The 
potential implications are discussed in \S5.

\section{Spectroscopy}

In order to test the nature of the giant star candidates, we obtained spectra
of twelve giant star candidates in CHVC 271+29+181 and its control field on 
UT January 27-28 2004 using the 6.5-m Clay telescope at Las Campanas with the MIKE spectrograph.  CHVC 271+29+181 
is one of the two fields in which both the HVC and control field have a clear excess of giant candidates beyond
the predicted contamination level.

The resolution of the echelle spectrograph is approximately $R\sim19,000$ in the red orders.  
To derive radial velocities, we used three orders extending from the atmospheric bands near 
8300 \AA\ to the Calcium IR triplet.
Details of the radial velocity measurement technique can be found in Majewski et al. (2004a).  In brief, 
radial velocities were derived using cross-correlation against a ``universal" template masked down
to a series of lines that confer the most power in the cross-correlation, and eliminating the 
vast portions of the spectrum that contain mostly continuum and therefore contribute little more
than noise in the cross-correlation.  Residual systematic offsets in the velocities were calibrated by
observations of K giant velocity standards.  The HVC radial velocities thus obtained 
were subsequently corrected for slit-centering errors by a second cross-correlation focused on the atmospheric 
absorption bands.  The resultant precision on the radial velocities from this multi-step process 
is better than 1 km s$^{-1}$.

Table 4 presents the heliocentric radial velocities for the brightest giant star candidates in CHVC 271+29+181
and its control field.  
Also listed in the table is the strength of each star's correlation
peak, a quality indicator from 0 (worst) to 7 (best) of the spectral quality as well as the de-reddened
$M$ magnitude and $(M-T_2)$ color. 

The most important aspect of the radial velocity distribution is the lack of stars at the radial velocity of 
CHVC 271+29+181 ($v_{r,LSR}$=+181 km s$^{-1}$; $v_{r,helio}$=+191 km s$^{-1}$).  The star nearest this 
radial velocity is 25 km s$^{-1}$ away and is likely to be a star from the Galactic stellar halo.  
{\it We believe that none of our spectroscopically observed giant candidates is associated with the HVC.}

\S3 and Table 3 indicate that approximately half of the giant candidates in CHVC 271+29+181 are Galactic
field stars (subdwarfs, stars with photometric errors and {\it bona fide} Galactic giants).  The origin
of the remaining half is unclear. 
However, our spectroscopic sample is bright and therefore expected to be heavily
dominated by Galactic field stars.  
The excess of giant candidates is more prevalent at fainter magnitudes.  

Figure 9 shows the radial velocity distribution of the stars and confirms 
that these are indeed Galactic field stars.  At the $(l,b)$ of CHVC 271+29+181 
the heliocentric radial velocities of any 
Galactic field stars will primarily reflect circular velocity differences.  One would expect roughly three
groupings of stars: a thin disk contribution near 0 km s$^{-1}$, a thick disk with a contribution near 
50 km s$^{-1}$ and a halo contribution near 220 km s$^{-1}$ (see, e.g. Casertano et al. 1990).  
Figure 9 shows that the radial velocity distribution is roughly consistent with this.  The spectroscopically 
observed stars lie
principally between -5 and 115 km s$^{-1}$ with a couple of stars at higher, halo-like radial velocities, 
including star 146, which is in retrograde rotation.

It is not surprising that these stars are Galactic field stars.
Inspection of Figure 7 shows that the brightest giants in these two fields are 
either very red or very blue, where the contamination from photometric error and subdwarfs is expected to 
be the worst.  Indeed, five of the six stars in the control field are at the red end and only slightly removed from 
the dwarf locus in the two-color diagram,
which would suggest that using a more rigid cut-off at the red end of the data may be more 
appropriate in the future.

Of course, the most interesting stars in CHVC 271+29+181 are the faint, moderately red stars that 
represent the true excess above the expected contamination level, particularly the
clump at ($M_0, (M-T_2)_0) \sim (21.5, 1.9$) in the control field.  Future spectroscopy of these stars 
will help to determine their nature.

\section{Discussion}

\subsection{Limits on the Stellar Content and Distance of the HVCs}

Previous searches for stellar populations in HVCs have concluded that there are not significant
populations of stars in HVCs.  Our survey of four HVCs here, meant to identify diffuse populations of giant stars, 
suggests that there may not even be {\it insignificant}
populations of stars in HVCs.  In \S3 and \S4 we show that the small number of giant candidates
we have identified along the line of sight to each of the HVCs we studied are likely to be field
stars from the standard Galactic stellar populations.
Our survey cannot, however, rule out the possibility of a stellar counterpart to the HVCs
that might be {\it displaced} from our probes.  Our survey probes consist of a single HVC and control field
pair for each target and, these imaged fields actually cover a small fraction of each HVC (see figure 1).
Were an HVC stellar population {\it both} spatially concentrated in its radial distribution {\it and} 
centered away from the center of the HI gas, we could erroneously infer a lack of stars.   

Ram-pressure stripping, tidal stripping and/or 
supernovae can potentially separate gas from stars and the halo abounds with examples.  The most 
famous high velocity
cloud -- the Magellanic stream -- is substantially separated from the stellar population of the LMC.
Gallart et al. (2001) argue for association of the Phoenix dwarf with an HI cloud that only 
partly overlaps the stellar populations.  The
Sculptor dSph has disconnected clumps of gas at similar velocities (Carignan et al. 1998), although this may 
be foreground contamination by the Magellanic stream (Putman et al. 2003).
It is possible that some of the excess stars in our data, particularly in the control fields, 
could represent a stellar population displaced from the center of the HI.  The
aforementioned clump of stars at ($M_0, (M-T_2)_0) \sim (21.5, 1.9$) in the control field of
CHVC 271+29+181 could be the RGB tip of a displaced stellar population.
Only further spectroscopy of the giant candidates can eliminate this possibility.

Although ours is the most sensitive probe so far for diffuse stellar populations, it is
important to be clear about the upper limits on HVC stellar populations that our null results place.
Though the distance, spatial distribution
and mass function of any hypothetical HVC stellar populations are unknown, a few basic 
assumptions allow us to derive useful order-of-magnitude limits.
We first define the distance to which we would be able to detect TRGB stars.
We estimated TRGB colors and magnitudes from the synthetic photometry of 
Ostheimer (2003) for a variety of old stellar populations.  The
TRGB $M-T_2$ color and $M_M$ absolute magnitudes are given in Table 5 for various metallicities.
Table 5 also lists minimum distances of a TRGB at which it would be fainter than $M_0=22$ and
$M_0=21.25$.\footnote[5]{We do not expect any of our giant candidates to be red supergiants given the lack 
of a young main sequence in the CMDs.
Post-Asymptotic Giant Branch supergiants could be produced by an old population but are short-lived 
and therefore rare (see Alves et al. 2001).}  

Beyond 550-760 kpc, RGB stars would be too faint for our program.  For an HVC at a hypothetical distance less
than these limits, however, we can make a rough analytical approximation of the HVC stellar mass implied by any
detected giant stars.  At distance $D$, the absolute magnitude of the TRGB ($M_{abs,TRGB}$) 
corresponds to an observed apparent magnitude 
$M_{app,TRGB} = M_{abs,TRGB} + 5 log_{10} \frac{D}{10 pc}$.
The limiting magnitude of our sample $M_{app,lim}$, in turn, corresponds to an absolute magnitude at distance
$D$ of $M_{abs,lim}=M_{app,lim} - 5 log_{10} \frac{D}{10 pc}$.  An estimate of the 
number of HVC giant candidates between $M_{app,TRGB}$ and $M_{app,lim}$ allows us
to integrate over the luminosity function and spatial distribution of the 
hypothetical HVC stellar population to estimate a total stellar mass.

The number of stars used for this exercise would optimally be taken from the excess of 
HVC field stars against control field stars.  However, the {\it deficit} of 
giant candidates in the HVCs compared with their control fields precludes this.
As noted above, however, the two CHVC fields and their control fields have approximately 
3 and 7 $\sigma$ more giant candidates than the predicted contamination level.
We therefore calculate implied stellar masses for these two CHVCs by taking the number
of giant stars as the lower of either the excess of giant candidates or the total
number of stars fainter than $M_{app,TRGB}$.  We also run calculations for all four HVCs
assuming that only {\it one} star in the entire sample represented the HVC stellar population.  The latter
calculation is a useful benchmark of the upper limit to the HVC stellar mass.

We integrate the number of giants over the complete global luminosity function of M3 
(Rood et al. 1999), assuming that any local group object would have an LF similar to that of 
globular clusters (see, e.g., Feltzing et al. 1999).  We convert the LF to Washington colors via 
the transformations in Majewski et al. 2000b), and assume an exponential spatial distribution with 
scalelength equal to the HWHM of the HI gas and axial ratio similar to that of the HI gas.  

Figure 10 shows the upper limit on stellar mass in the HVCs plotted against distance.
The solid lines show the fundamental limitation of our survey.  A single HVC star in our sample would imply 
stellar masses of a few $10^5 M_{\sun}$.  CHVC 017-25-218 is close to its fundamental limit, owing to the 
comparatively small size of the HVC and the brightness of its excess giant candidates.  CHVC 271+29+181 has a 
higher potential stellar mass ($10^6$ $M_{\sun}$).  However, this higher mass limit is
a result of the large number of faint giant candidates in and large spatial extent of the HVC .  
We emphasize that these
are upper limits and that the CHVC stellar content is likely well below this limit, if it exists at all.  

\subsection{Halo Streams?}

Our predicted giant star candidate counts match the observed counts reasonably well (see table 3).  However, 
the two compact HVCs show an interesting statistical signature.  
CHVC 017-25-218 and its control field both have approximately
3$\sigma$ more giant candidates than predicted while CHVC 271+29+181 and its control field have 3$\sigma$ and
7$\sigma$ more giant candidates, respectively.

{\it It is unlikely that these stars are associated with the clouds themselves}, as show in the lack of excess
in the CHVC central fields over the controls.  However, 
these stars may still be interesting objects in their own right.
Our interest is particularly drawn to the clump of stars at ($M_0, (M-T_2)_0) \sim (21.5, 1.9$) in the 
control field of CHVC 271+29+191, which could be the tip of a very distant RGB.

One important caveat in the Galactic model used in \S3.4, as noted in Siegel et al. (2002), 
is that the halo often defies conventional density laws.  Siegel et al. argued for two halo populations --
a smooth inner halo (which we model) and a highly structured outer halo (which we can't and don't), 
possibly comprised of coherent streams of stellar debris from disrupted dwarf galaxies and/or 
globular clusters.

Several recent studies using our Washington system have found groups of 
distant giant stars that are not clearly associated with any known dwarf galaxies or globular clusters.  
Mu\~noz et al. (2004) have recently identified a velocity coherent group of halo giants in the line of sight
of the Carina dSph galaxy, but which are not associated with the core of that object.
In a study around the Magellanic Clouds, 
Majewski et al. (in preparation; see also Majewski 2004) 
show that, at large Galactocentric distances, the Galactic 
stellar halo may be {\it dominated} by coherent
streams of stars, presumably tidally stripped by the Milky Way from dwarf galaxies or globular
clusters.  The Majewski et al. fields do {\it not} show clear RGB structure in the CMD because the streams 
are diffuse -- a result eerily similar to what we have found in our two CHVC fields.  Additionally, Ostheimer's 
(2002) deep Washington photometry survey of M31 uncovered numerous examples of potentially {\it intergalactic} 
giant star candidates, as well as one coherent halo substructure at a distance of about 20 kpc 
(Majewski et al. 2004b), now confirmed by a strong MSTO feature in his survey data.
Interestingly, even this latter, ``Triangulum-Andromeda" structure does not exhibit a strongly identifiable
RGB in the CMD, but one that is more subtle (Rocha-Pinto et al. 2004).
These precedents make it possible to infer that the excesses in the HVC
fields may represent distant diffuse star streams in the outer halo that just happen to cross our survey fields.

CHVC 017-25-218 is particularly interesting because its candidate giants are bright.  
While the HVC and control fields each have
approximately twenty giant candidates fainter than $M\sim19$ --- a number consistent with subdwarf, extragalactic
and photometric contamination --- each of these fields also has approximately forty giant candidates 
brighter than $M\sim19$, which is nearly five times the expected contribution from the canonical Galactic models.
This field star overdensity could be a signature of halo substructure in the field.

CHVC 017-25-218 is close to the extended stellar stream of the tidally disrupted Sagittarius
dSph galaxy (Majewski et al. 2003) and the distance modulus of the Sagittarius stream near this location ($m-M\sim
18$) would put Sgr RGB stars close to the apparent magnitude of the bright clump in CHVC 017-25-218 and its 
control field.  It is highly probably that the excess of bright giants in CHVC 017-25-218 and its control field are 
giants
from the Sagittarius debris stream, although spectroscopy will be needed to confirm this interpretation.  If these
stars are Sagittarius debris, it validates the basic strategy of our survey since in this case we {\it would} have
successfully identified a diffuse,
but distinct stellar population in our fields.  It should be noted that even if the stars are Sagittarius debris, this 
says nothing about the CHVC itself.  The radial velocity of CHVC 017-25-218 is a few hundred km s$^{-1}$ discrepant 
from the Sgr radial velocity near this field.

In \S4, we showed that the brighter giants in CHVC 271+29+181 and its control field are Galactic field stars. 
However, the bulk of the giant candidates are fainter than the limit of our spectroscopy (and fainter than the excess 
in CHVC 017-25-218 and its control field).  It is possible that spectroscopy of the fainter giants ---
particularly the ($M_0, (M-T_2)_0) \sim (21.5, 1.9$) group in the control field --- might
reveal another moving group of giant stars even further into the halo.

\section{Summary}

Our multifilter examination of four HVCs has failed to detect evidence for a 
stellar population in four HVCs, which, by virtue of our search strategy for diffuse populations
of giant stars, places the strongest limits so far
on the existence of stellar populations in the cores of HVCs.
Although a number of stars in our survey fields can be classified as distant giants based on their location in
two-color space, there is no excess of giant candidates
in any HVC compared with its control field.  In addition, the total number of giant candidates in our fields
roughly corresponds to the predicted contribution from Galactic field stars.
Echelle spectroscopy of the brightest giants in one
HVC and its control field confirm that they are indeed likely to represent stars from canonical Galactic populations.

The two compact HVCs show a significant excess of giant candidates beyond the predicted contamination level.  However,
it is likely that these excesses are unrelated to the CHVCs and are, in fact, distant   
overdensities (perhaps streams) of halo giants unrelated to
the HVCs themselves, particularly in CHVC 017-25-218, which nearly overlaps the Sagittarius debris stream.
Only further spectroscopy of the giant candidates should definitively resolve these issues.

If none of our stars are associated with their component HVCs, this places an upper limit of a 
few $10^5 M_{\sun}$ on the stellar content of the HVCs out to a distance of 0.6 Mpc.

\acknowledgements

The authors would like to thank J.D. Crane and R.J. Patterson for taking the Mosaic data,
J. Rhee and J. Ostheimer for sharing their synthetic photometry
for TRGB absolute magnitude calibrations and A. Bernacchi for help reducing the MOSAIC data.
This manuscript was much improved by the patient and deligent work of the anonymous referee.
The authors also thank M. Putman for helpful discussions.  Support for this program was provided to SRM by National
Science Foundation (NSF) CAREER Award AST-9702521, NSF grant AST-0307851,
the David and Lucile Packard Foundation, a Cottrell Scholar Award from the Research Corporation and a Space 
Interferometry Mission Key Project grant, NASA/JPL contract 1228235.
CG was partially supported by the Spanish Ministry of Science
and Technology (Plan Nacional de Investigacion Cientfica, Desarrollo e
Investigacion Tecnolgica, AYA2002-01939), and by the European Structural Funds.

\clearpage

{\bf Figure Captions}

\figcaption[f1a.eps]{HI column density distribution in the four HVC fields. The HI
emission at about $15\arcmin$ resolution as observed with HIPASS (Barnes
et al. 2001, Putman et al. 2002) from a 7$\times$7 degree field integrated
over the velocity extent of each object and scaled to column density
assuming negligible HI opacity. The greyscale and contour values are in
units of 10$^{18}$cm$^{-2}$. The greyscale ranges are indicated by the
labeled wedge at the top of each plot. Contours are drawn at 1, 2, 5, 10,
20, 50 and 100 $\times$ 10$^{18}$cm$^{-2}$. The overlaid boxes indicate
the positions and approximate sizes of the central and control fields
observed for each object.}

\figcaption[f2.eps]{Magnitude error as a function of magnitude for all our passbands.  DuPont
data are on the left, Mayall data on the right.  The horizontal lines mark our error limit while the
vertical dashed lines indicate the $M$ magnitude limits.  The larger uncertainties for the 2.5-meter data reflect
the poorer PSF of the WFC more than the smaller aperture.}

\figcaption[f3.eps]{Color-magnitude diagrams of (left, top to bottom) HVC 017-25-218,
030-51-119, 267+26+216, and 271+29+181, along with their corresponding
control fields (right, top to bottom).  Note the lack of any 
obvious differences between the HVC and control fields.  Note also that any
sparse RGB would be swamped by the sheer number of stars in each field.}

\figcaption[f4.eps]{Color-color diagrams of the four HVC fields and control
fields.  Panels are as in Figure 3.  Note the clear separation of the stars into
two populations beyond $M-T\sim1$ -- the extended ``swoosh'' of high-surface gravity dwarfs 
and the sprinkling of low-surface gravity giant candidates.}

\figcaption[f5.eps]{Giant star selection.  The plots show $M$ magnitude against $\Delta$ -- a measure
of the separation of each star from the dwarf locus.  Star are selected as giants (large dots) if they 
are more than 3$\sigma$ away from the mean $\Delta$ at that magnitude.  Starred points are those with
$\Delta \leq -3\sigma$ and provide a measure of the photometric contamination level.}

\figcaption[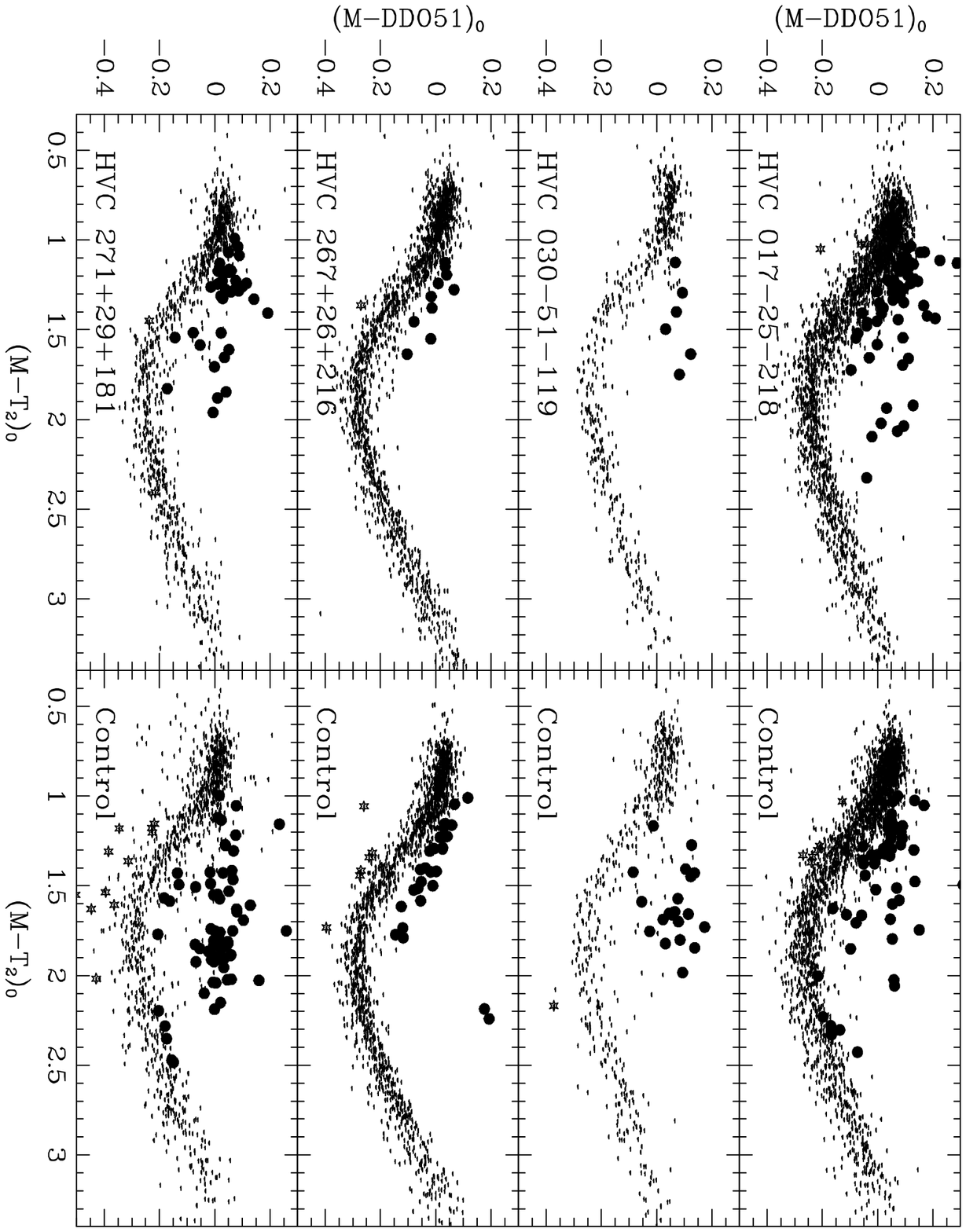]{Color-color diagrams of the four HVC fields and control
fields.  Panels are as in Figure 2.  Large dots are objects selected as potential giant stars based
on their $\Delta$ values while starred points have $\Delta \leq -3\sigma$ and provide a measure of the photometric contamination level.
HVC 267+26+216 shows far less scatter in the two-color diagram because of the shallower limiting
magnitude and the superior imaging of the 4-meter telescope.}

\figcaption[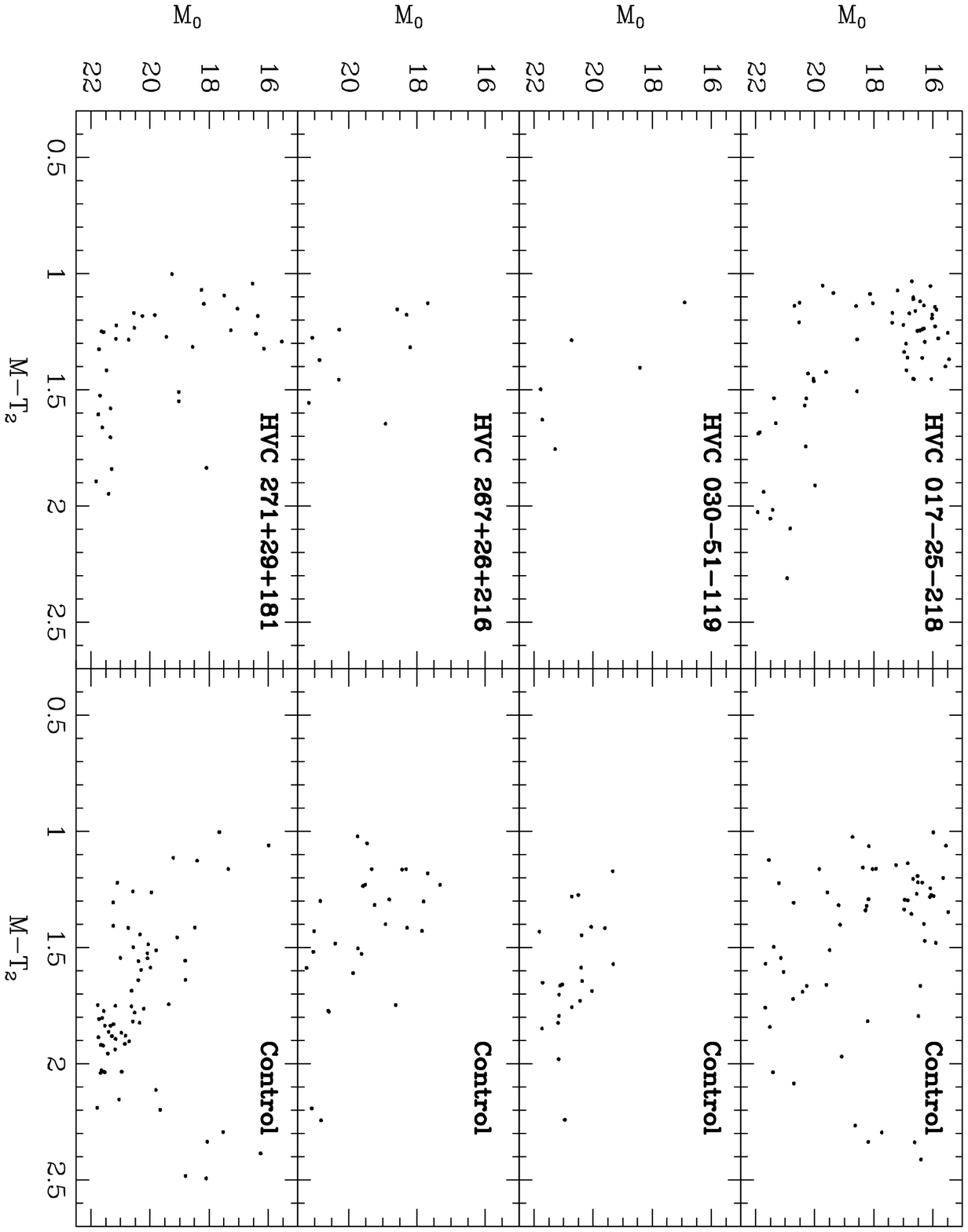]{Color-magnitude diagrams of giant candidates selected from
the four HVC fields and their control fields.  Panels are as in Figure 2.  Note
the lack of any obvious RGB in the distribution of giant candidates.  Note also 
that most of the candidates are at the faint end of the data, with the exception of
HVC 017-225-218.}

\figcaption[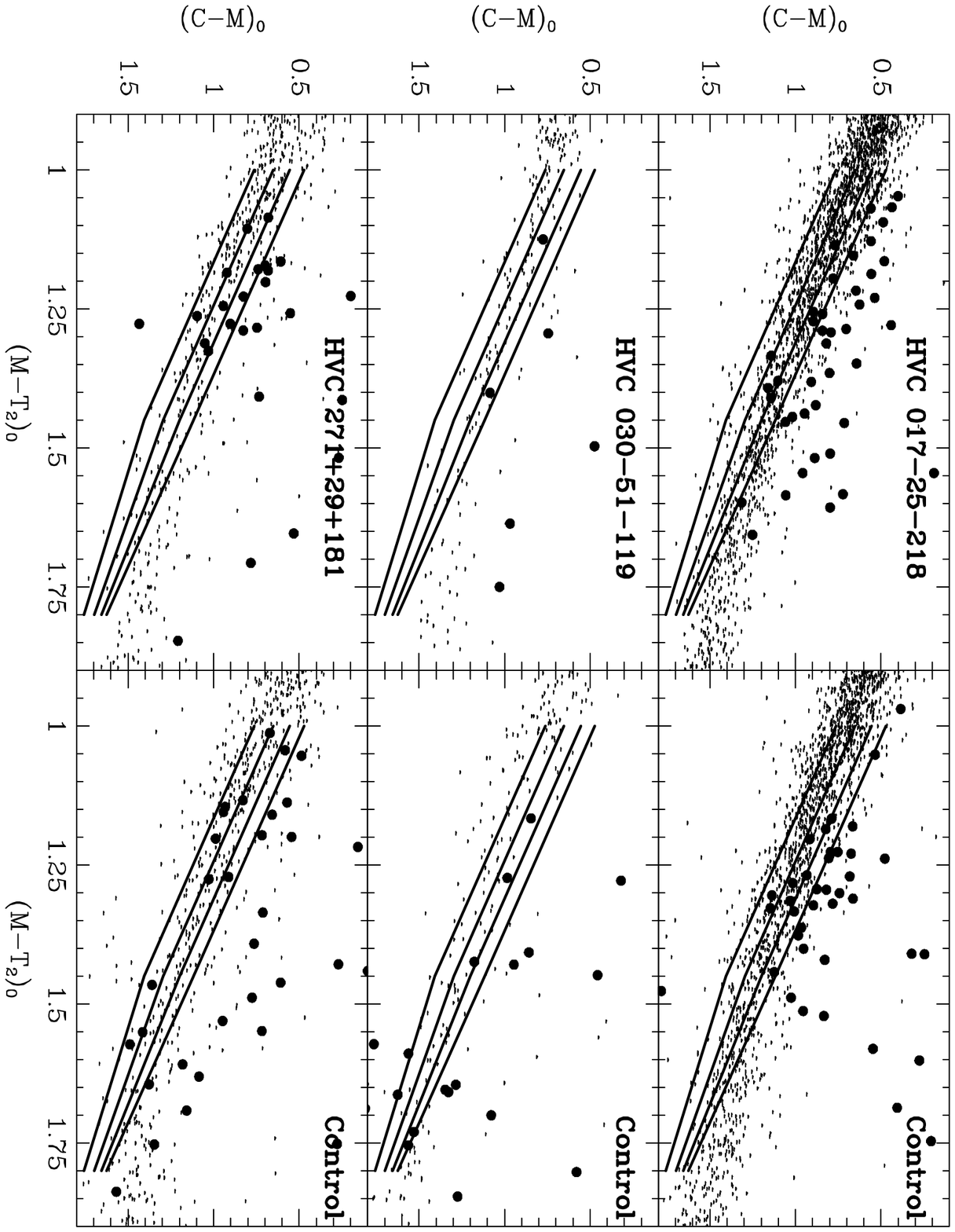]{($C-M$, $M-T_2$) two color diagrams of giant candidates (filled circles) with 
$\sigma_C \le 0.08$ from the HVCs and control fields.  The dots are dwarf stars.  The solids lines 
are iso-metallicity lines from Geisler et al. 1991.  These lines
only extend to $M-T_2=1.8$.  The lines represent (bottom to top) [Fe/H] = -0.5, -1.0, -1.5 and -2.0.  
Note that this two-color diagram compresses surface gravity information.  Both 
dwarf and  giant stars will lie along these lines.  The gaps between the lines narrows at red colors because the $C$ filter becomes 
less sensitive to abundance at low temperatures.  Many of our giant candidates are well blueward of 
these lines, indicating that they are not stellar objects.  These two-color diagrams can be 
contrasted with the coherent RGBs shown in Figure 7 of Geisler et al..  The paucity of stars 
in HVC 030-51-119 and its control field reflects the minimal 
data obtained in the $C$ filter for those fields.}

\figcaption[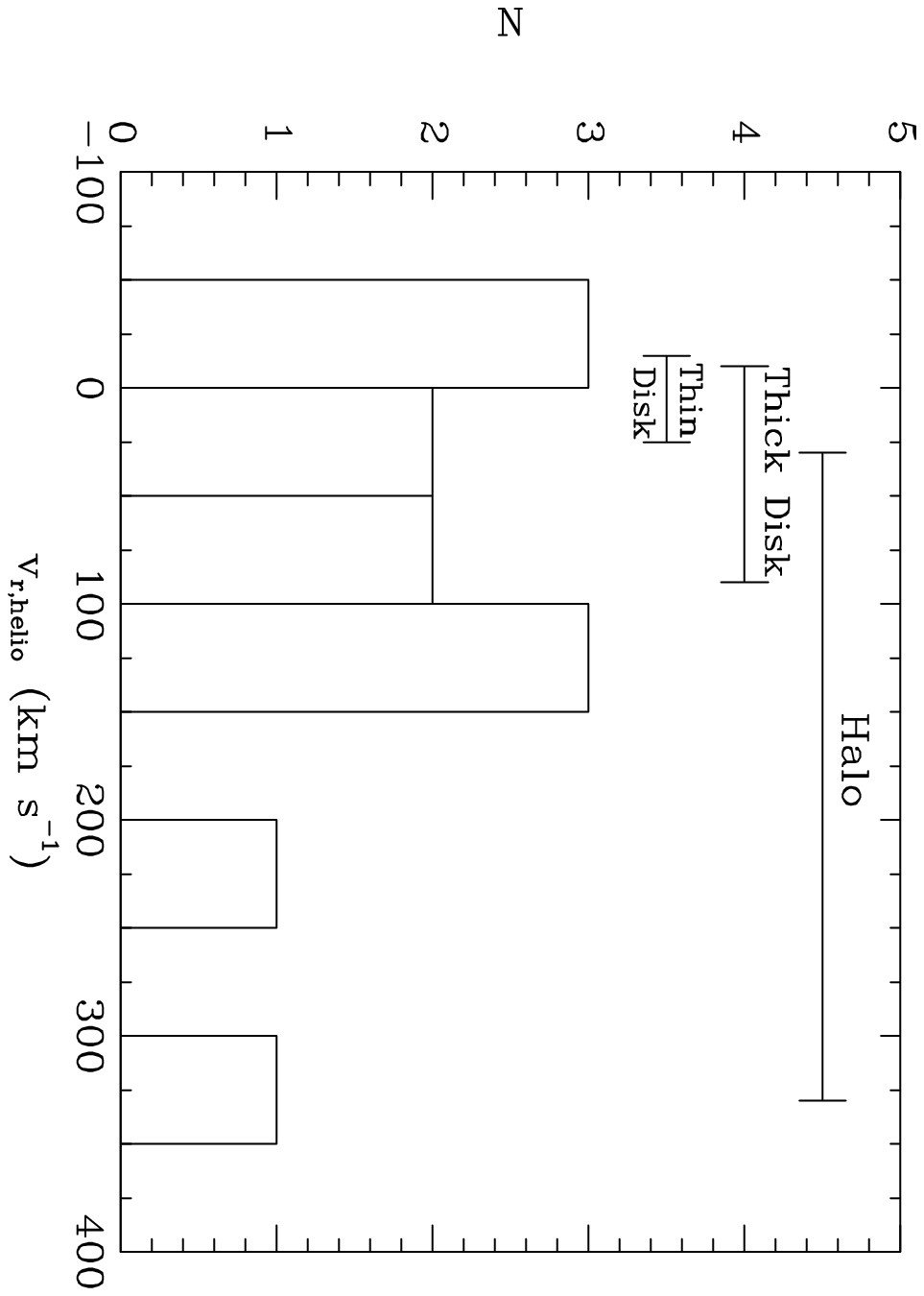]{The heliocentric radial velocity distribution of stars in CHVC 271+29+181 and its 
control field.  Overlayed are the approximate radial velocity ranges expected for the canonical thin
disk, thick disk and halo velocity distributions from Casertano et al. (1990).
Note that the data are roughly consistent with the canonical
Galactic stellar populations.  Note also the dearth of stars near +191 km s$^{-1}$, the heliocentric radial
velocity of the CHVC.}

\figcaption[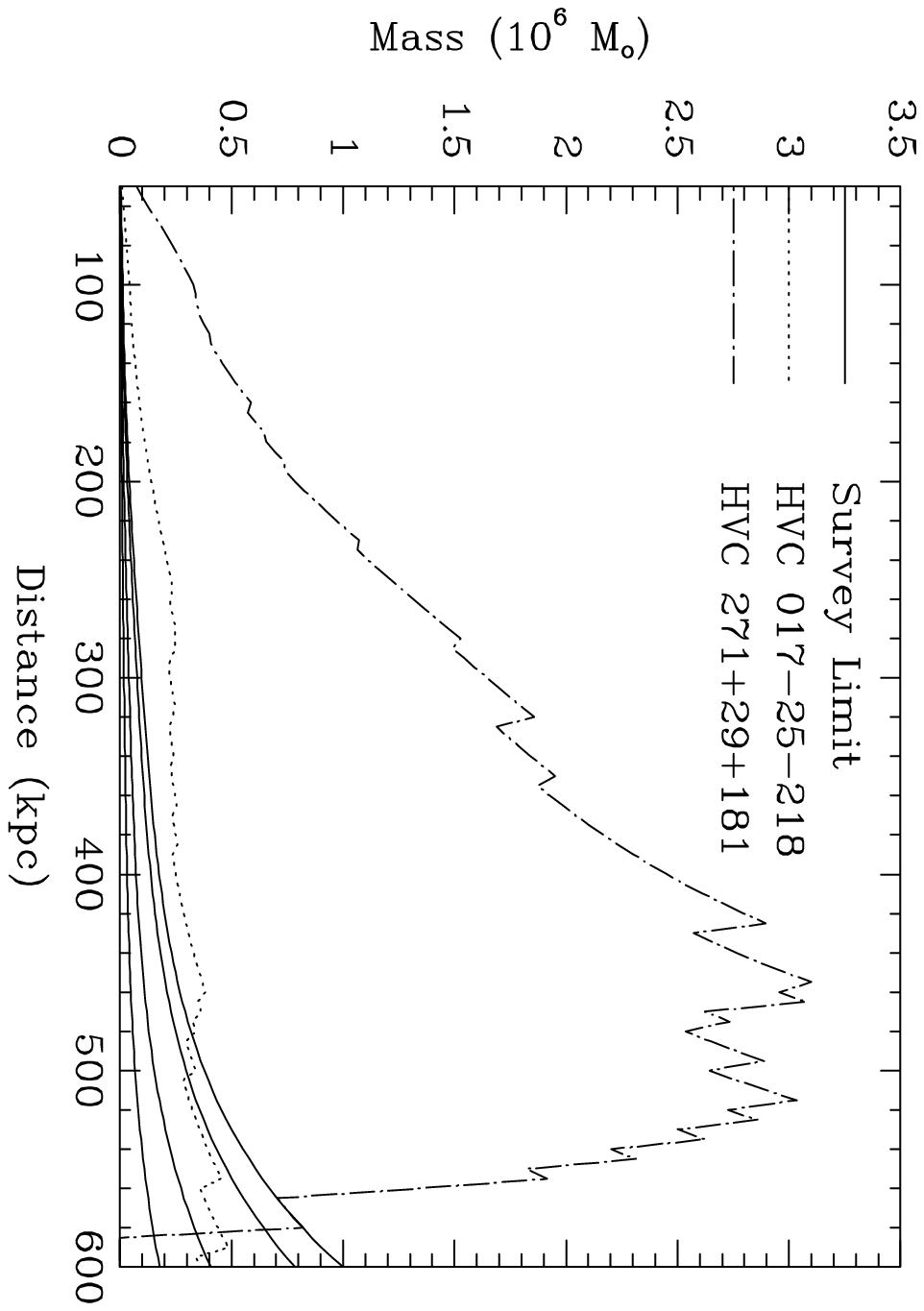]{The upper limit of stellar mass in each of our four HVCs as a
function of distance along the line of sight.  The rollover at large distances is
the magnitude limit of the data.}

\clearpage

\begin{deluxetable}{cccc}
\tablewidth{0 pt}
\tablecaption{Target Fields}
\tablehead{
\colhead{Field} &
\colhead{$(\alpha,\delta)_{2000.0}$} &
\colhead{$(l,b)$} &
\colhead{$E_{B-V}$}}
\startdata
CHVC 017-25-218    & 19:59:06.0,-24:55:00 & 16.6,-25.3 & 0.107\nl
Control 017-25-218 & 20:00:18.5,-23:04:03 & 18.6,-24.9 & 0.148\nl
\hline
HVC 030-51-119    & 21:58:18.0,-22:42:00 & 29.6,-50.7 & 0.044\nl
Control 030-51-119 & 21:59:55.9,-21:24:01 & 31.7,-50.7 & 0.037\nl 
\hline
HVC 267+26+216    & 10:28:06.15,-26:40:44.0 & 267.2,26.2 & 0.077\nl
Control 267+26+216  & 10:36:26.18,-27:56:49.5 & 270.5,26.2 & 0.061\nl
\hline
CHVC 271+29+181    & 10:48:54.0,-26:23:00 & 271.5,28.9 & 0.051\nl
Control 271+29+181 & 10:57:22.2,-27:29:12 & 274.0,28.9 & 0.068\nl
\enddata
\end{deluxetable}

\begin{deluxetable}{lcccc}
\tablewidth{0 pt}
\tablecaption{Number Counts}
\tablehead{
\colhead{Field} &
\colhead{$N_{stars,HVC}$} &
\colhead{$N_{stars,Control}$} &
\colhead{$N_{giants,HVC}$} &
\colhead{$N_{giants,Control}$}}
\startdata
CHVC 017-25-218 & 3179 & 2973 & 63 & 60 \\
HVC 030-51-119  &  555 &  550 &  6 & 22 \\
HVC 267+26+216  & 1867 & 1885 & 10 & 28 \\
CHVC 271+29+181 & 1032 & 1262 & 35 & 67 \\
\enddata
\end{deluxetable}

\begin{deluxetable}{lccccc}
\tablewidth{0 pt}
\tablecaption{Giant Region Contamination}
\tablehead{
\colhead{Field} &
\colhead{$N_{expected}$} &
\colhead{$N_{\Delta \geq 3\sigma}$} &
\colhead{$N_{Galactic}$} &
\colhead{$N_{Total}$} &
\colhead{$N_{Excess}$}}
\startdata
CHVC 017-25-218    & 10 & 4  & 8 & 43 & 20\\
Control 017-25-218 &  9 & 6  & 8 & 42 & 18\\
\hline
HVC 030-51-119     &  2 & 0  & 2 & 19 & -13\\
Control 030-51-119 &  2 & 3  & 2 & 20 & -2\\
\hline
HVC 267+26+216     &  6 & 1  & 6 & 27 & -17\\
Control 267+26+216 &  6 & 6  & 7 & 28 & 0\\
\hline
CHVC 271+29+181    &  3 & 2  & 2 & 20 & 15\\
Control 271+29+181 &  4 & 11 & 2 & 28 & 39\\
\enddata
\end{deluxetable}

\begin{deluxetable}{lcrcccc}
\tablewidth{0 pt}
\tablecaption{Giant Candidate Radial Velocities}
\tablehead{
\colhead{Field} &
\colhead{Star ID} &
\colhead{$v_{r,helio}$} &
\colhead{Correlation} &
\colhead{Q} &
\colhead{$M_0$} &
\colhead{$(M-T_2)_0$} \\
\colhead{} &
\colhead{} &
\colhead{km s$^{-1}$} &
\colhead{Peak} &
\colhead{(0-7)} &
\colhead{} &
\colhead{}}
\startdata
CHVC 271+29+181    & 240 & 147.01 & 0.67 & 4  & 17.49 & 1.09 \\
CHVC 271+29+181    & 231 &  85.21 & 1.01 & 7  & 17.27 & 1.24 \\
CHVC 271+29+181    & 206 &  40.55 & 0.71 & 6  & 17.04 & 1.15 \\
CHVC 271+29+181    & 160 & 115.01 & 1.10 & 7  & 16.41 & 1.26 \\
CHVC 271+29+181    & 146 & 349.69 & 0.82 & 7  & 16.53 & 1.04 \\
CHVC 271+29+181    & 142 &  66.76 & 0.65 & 7  & 16.36 & 1.18 \\
CHVC 271+29+181    & 126 & 106.95 & 0.84 & 7  & 16.15 & 1.32 \\
CHVC 271+29+181    &  87 & -44.42 & 1.10 & 7  & 15.54 & 1.29 \\
Control 271+29+181 & 517 &  24.67 & 0.66 & 6  & 18.10 & 2.49 \\
Control 271+29+181 & 445 &  -8.33 & 0.83 & 7  & 17.52 & 2.29 \\
Control 271+29+181 & 134 & 215.08 & 0.79 & 7  & 15.99 & 1.06 \\
Control 271+29+181 & 93  &  -4.63 & 1.15 & 7  & 16.26 & 2.38 \\
\enddata
\end{deluxetable}

\begin{deluxetable}{cccccc}
\tablewidth{0 pt}
\tablecaption{HVC Distance Limits}
\tablehead{
\colhead{[Fe/H]} &
\colhead{Age} &
\colhead{$(M-T_2)_{TRGB}$} &
\colhead{$M_{M,TRGB}$} &
\colhead{$r_{\sun,M=22}$} &
\colhead{$r_{\sun,M=21.25}$}\\
\colhead{} &
\colhead{(Gyr)} &
\colhead{} &
\colhead{} &
\colhead{(kpc)} &
\colhead{(kpc)}}
\startdata
-2.0 & 10-15 & 1.78 & -2.41 & 760 & 540\nl
-1.5 & 10-15 & 1.94 & -2.28 & 720 & 510\nl
-1.0 & 10-15 & 2.22 & -1.99 & 630 & 450\nl
-0.5 & 10-15 & 2.49 & -1.70 & 550 & 390\nl
\enddata
\end{deluxetable}

\clearpage
%\begin{figure}[h]
%\plotone{f1a.jpg}
%\end{figure}
%\begin{figure}[h]
%\plotone{f1b.jpg}
%\end{figure}
%\begin{figure}[h]
%\plotone{f1c.jpg}
%\end{figure}
%\begin{figure}[h]
%\plotone{f1d.jpg}
%\end{figure}
%\clearpage
%\begin{figure}[h]
%\plotone{f2.jpg}
%\end{figure}
%\begin{figure}[h]
%\plotone{f3.jpg}
%\end{figure}
%\begin{figure}[h]
%\plotone{f4.jpg}
%\end{figure}
%\begin{figure}[h]
%\plotone{f5.jpg}
%\end{figure}
%\clearpage
\begin{figure}[h]
\plotone{f6.eps}
\end{figure}
\begin{figure}[h]
\plotone{f7.eps}
\end{figure}
\begin{figure}[h]
\plotone{f8.eps}
\end{figure}
\begin{figure}[h]
\plotone{f9.eps}
\end{figure}
\begin{figure}[h]
\plotone{f10.eps}
\end{figure}

\end{document}